# Genomic Materials Design: CALculation of PHAse *Dynamics*

G. B Olson [1,2] and Z. K. Liu [3]

[1] Department of Materials Science and Engineering, Massachusetts Institute of Technology,

Cambridge, MA 02139

[2]QuesTek Innovations LLC,

Evanston IL 60201

[3]Department of Materials Science and Engineering, Pennsylvania State University,

University Park, PA 16803

Emails: gbolson@mit.edu, prof.zikui.liu@psu.edu

Abstract

The CALPHAD system of fundamental phase-level databases, now known as the Materials Genome, has enabled a mature technology of computational materials design and qualification that has already met the acceleration goals of the national Materials Genome Initiative. As first commercialized by QuesTek Innovations, the methodology combines efficient genomic-level parametric design of new material composition and process specifications with multidisciplinary simulation-based forecasting of manufacturing variation, integrating efficient uncertainty management. Recent projects demonstrated under the multi-institutional CHiMaD Design Center notably include novel alloys designed specifically for additive manufacturing. With the proven success of the CALPHAD-based Materials Genome technology, current university research emphasizes new methodologies for affordable accelerated expansion of more accurate CALPHAD databases. Rapid adoption of these new capabilities by US apex corporations has compressed the materials design and development cycle to under 2 years, enabling a new

“materials concurrency” integrated into a new level of concurrent engineering supporting an unprecedented level of manufacturing innovation.

## 1 Introduction

As John Agren has pointed out [1], the real context of CALPHAD technology is what began as the US national Materials Genome Initiative (MGI) announced by President Obama in 2011 which has now become a global activity. The overarching goal of this is to develop and promulgate the technology that can take what has historically been a 10 to 20 year materials development cycle and compress that by at least 50 percent. As Agren mentioned, the original US National Academy study that called for this initiative was back in 2004 on the very topic of accelerating the transition of materials technology [2]. It looked at the analog of the Human Genome Project (HGP), acknowledging that it is the nature of the human genome to be an actual physical database that directs the assembly of the dynamic structures of life. The concept in 2004 was that an equivalent genome for materials would be the fundamental parameters that direct the assembly of the dynamic microstructures of materials. Inspired by the success of HGP and CALPHAD-based systems materials design established by Olson [3], Liu, one of the authors of the present paper, introduced the term "Materials Genome"® in 2002 [4] when incorporating his company Materials Genome, Inc. and starting the integration of first-principles calculations based on density functional theory (DFT) [5], shortly after he became the Editor-In-Chief of the CALPHAD Journal. Liu then trademarked it and agreed for its use in the MGI by the US government.

The present manuscript is based on the lead author's invited presentation at the CALPHAD Global 2021 virtual conference [6] and is collected as part of the special issue of the conference. The second author was invited to expand the manuscript by adding the perspectives on the advancements in databases and computational tools for the CALPHAD approach.

## 2 CALPHAD as Materials Genome

An historic milestone in the development of this technology that just predates MGI was the first flight of aircraft landing gear of QuesTek Ferrium S53 stainless landing gear steel [7]. It was the first stainless steel to meet the mechanical performance requirements of landing gear allowing the elimination of toxic cadmium plating, but much more significantly, it was the first fully computationally designed and qualified material to make it all the way to flight. This really does measure a level of maturity of this technology and that is reinforced by the timeline of Fig. 1 [8]. It is clear that the genome we have is the CALPHAD system of fundamental phase level data. While any nucleation event has many precursor fluctuations, if we follow the thread of the evolution of CALPHAD, we can trace its origins back to Larry Kaufman and Morris Cohen in 1956 [9]with their analysis of the iron-nickel binary system. In the way that we teach thermodynamics, there is a common misconception that thermodynamics is about equilibrium. If that were the case we would have called it "thermostatics". Thermodynamics was actually created to describe the dynamics of heat engines where equilibrium calculations set useful bounds on the real dynamic problems of interest. The same is true for the problems we address in materials engineering where it is the dynamics that we created these tools to deal with. The origin of the CALPHAD acronym coined by Kaufman was to counter the term PHACOMP in which there was a practice of trying to estimate solubility limits from the attributes of a single phase. Of course, Kaufman wanted to acknowledge that those limits really come from phase competition as expressed by phase diagrams and hence the CALPHAD acronym.

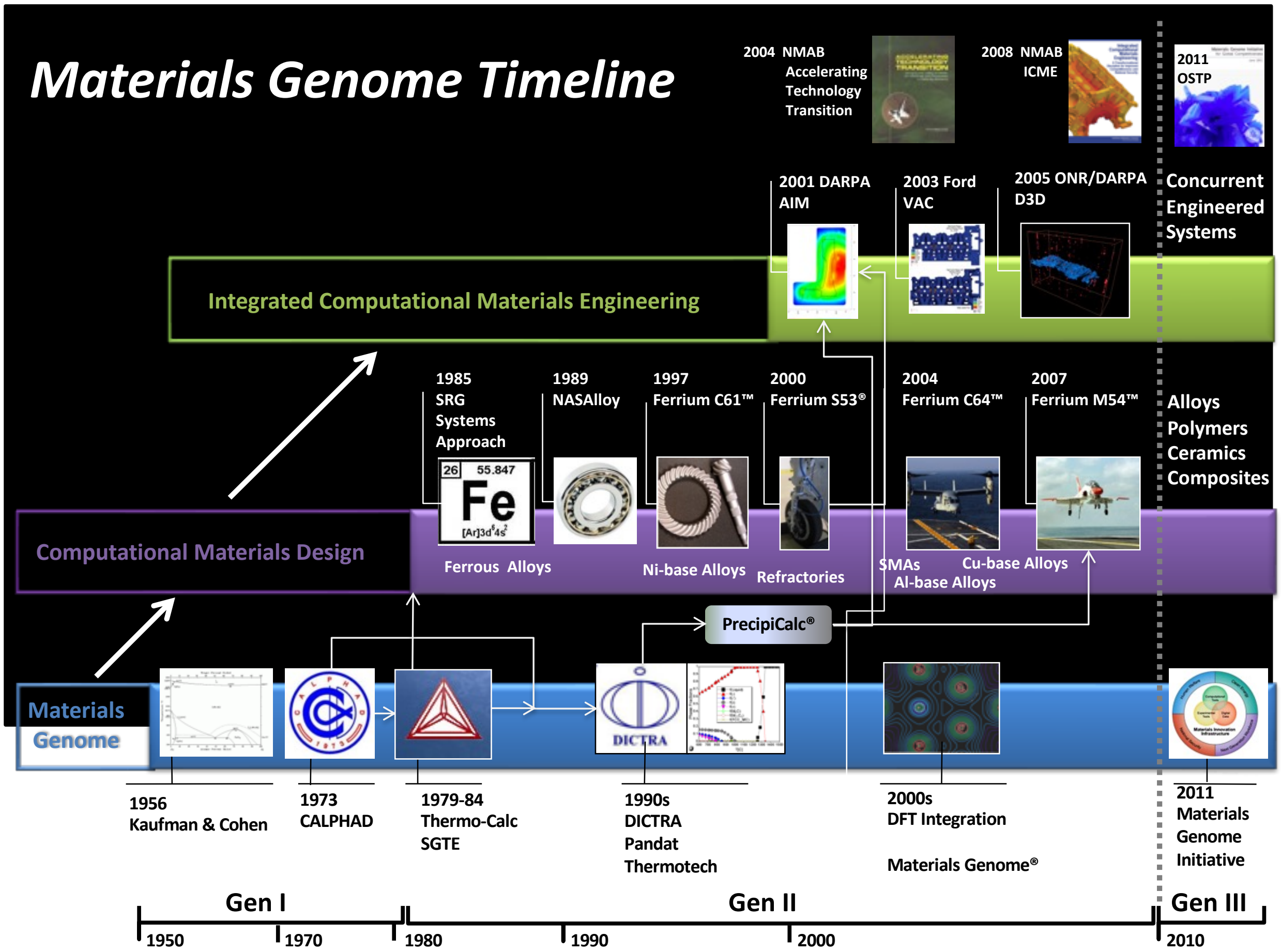

*Fig. 1: Time evolution of CALPHAD-genomic materials technology spanning 3 levels: (a) CALPHAD genomic level databases and software, (b) computational materials design, and (c) simulation-based accelerated materials qualification* [8].

However, if we look at what Larry Kaufman and Morris Cohen were actually doing with the iron-nickel system it had nothing to do with calculating phase diagrams. It was taking the information in an equilibrium phase diagram and reducing it to its more fundamental underlying thermodynamics, specifically to treat systems far from equilibrium, in their case martensitic transformations. For that purpose of treating the dynamics of martensite, it was Morris Cohen's vision that by having accurate low temperature thermodynamics that appropriately acknowledged the third law curvature of free energies, studying the time dependent low temperature transformation defines the dependence of activation energy on driving force that would help to understand the actual governing mechanism of martensitic transformations. That did come to pass over time, and we now have a very predictive theory of martensitic transformations [10] in which that first CALPHAD exercise in the 1950s set the groundwork.

## 3 CALPHAD-Genomic Materials Design

### 3.1 Overview and Design Principles

When Olson, the lead author of the present paper, started to work with Kaufman in 1970 in the research group of Morris Cohen at MIT, it was clear that Kaufman was much more an engineer than a scientist, but he was committed to science-based engineering; really very visionary at the time. His company Manlabs was a metallurgical consulting company that offered CALPHAD services, and to a large extent was a precursor of what QuesTek is today. We have a lot to thank Kaufman for. The evolution of CALPHAD very quickly moved beyond solution thermodynamics to include mobilities to deal with dynamic problems and has over time expanded not only in the scope of the systems it describes but the various phase level attributes that are described within the framework. Again, it is the control of dynamic systems that motivate that expansion of those parameters. It was the appearance of Thermo-Calc as a commercial software system and supporting database system that inspired our founding in 1985 of the "SRG" design consortium to build a methodology of computational materials design intended to be general. The activity first focused on high-performance steels, acknowledging this class of materials as the one we had studied the longest and deepest and had the highest quality of fundamental data at the time.

The real function of CALPHAD in this context of design is that it allows us to apply the mechanistic fundamental understanding we already possess, but in a system-specific quantitative form to enable a science-based form of engineering. The first steel designs ultimately led to the founding of QuesTek in the late 1990s. Meanwhile at the university, various demonstration projects across different classes of materials using the same methodology were performed. It was the successes in computational materials design demonstrated in the 1990s that built the case for the DARPA AIM initiative in the early 2000s which was the start of what we now call ICME [2,11]. This was to go beyond the initial design of material to address the full materials development cycle, which is the current goal of MGI. The DARPA AIM method linked microstructural models to macroscopic process simulators to accelerate process optimization at the component level. Very importantly, it simulated manufacturing variation within allowed tolerances to give accurate forecasts of minimum properties that become the design allowables that a material user can depend on. This was to replace statistical inference with probabilistic science that could greatly reduce the amount of experimentation necessary to bring a material all the way to certification.

Today we have over 60 years of our fundamental CALPHAD genome, over 30 years of a design practice, and 20 years of a full ICME technology with the cycle compression. This is allowing for the first time the expansion of concurrent engineering that used to mean everything but materials to now include materials in the concurrent engineering cycle with a profound impact on what concurrent engineering can achieve today.

The core of the approach comes from the philosophy of the late Cyril Stanley Smith who wrote about the intrinsic complexity of multi-scale dynamic materials microstructure, therefore advocating that we should take the systems approach that is used in other branches of engineering as the framework to deal with material complexity. In taking that to heart, a core concept was the linearization of the interaction of the four domains of material science of engineering advocated by Morris Cohen. He described it as a “reciprocity” between the opposite philosophies of science and engineering where the deductive cause and effect logic of predictive science moves from left to right and the inductive logic of the goal-means relations of engineering moves from right to left as shown in Fig. 2 [8]. We’ve used that as the backbone of a system structure of materials adding it to Smith’s structural hierarchy (see Fig. 3).

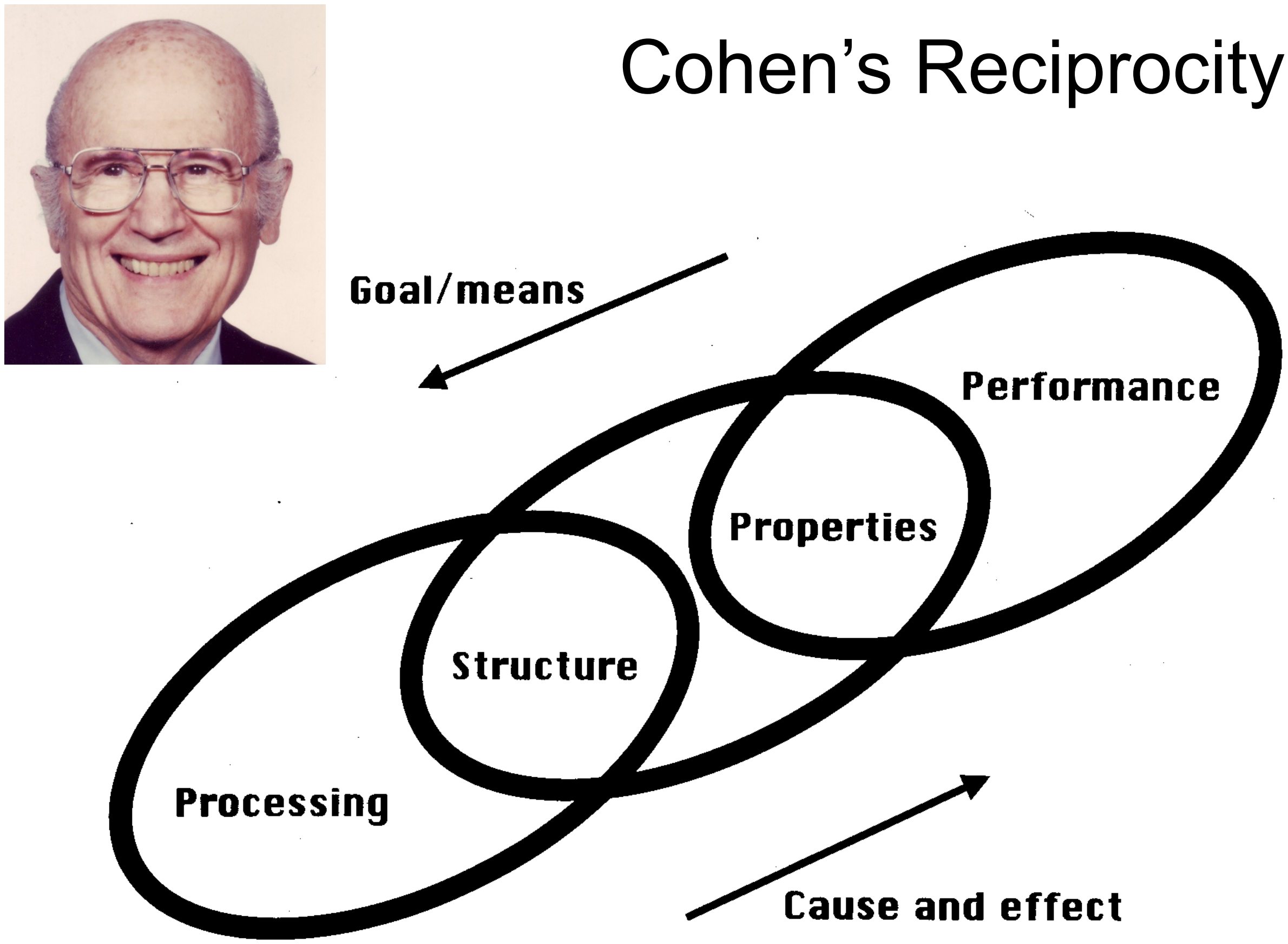

*Fig. 2: Three-link chain structure representing Morris Cohen's "reciprocity" between the opposite philosophies of science and engineering* [8].

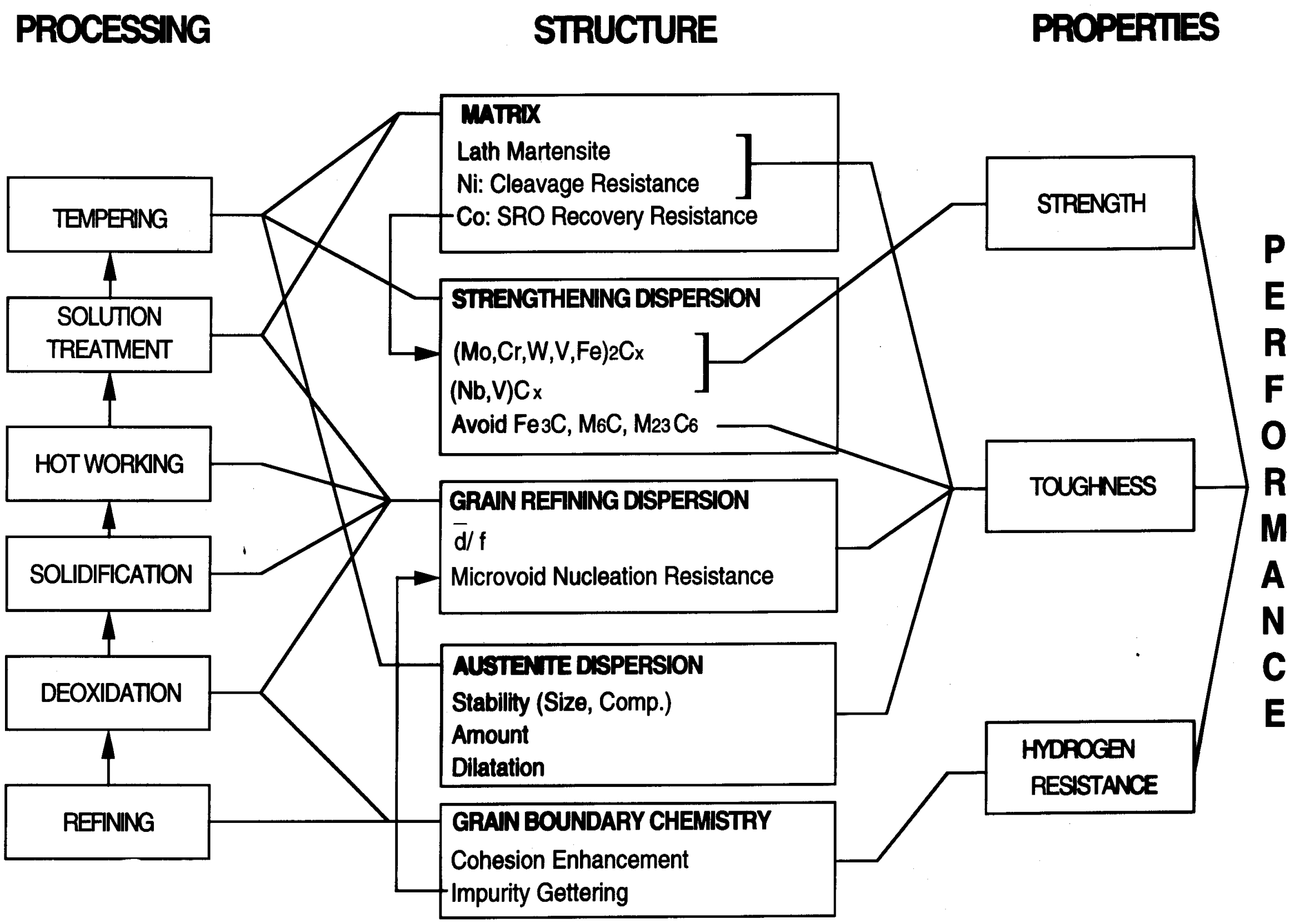

*Fig. 3: Hierarchical system structure of secondary hardening Ni-Co martensitic steel* [8].

Each design project starts with this kind of system chart where a performance goal of the material can be mapped to a quantitative set of property objectives which can be related back to subsystems of hierarchical material structure which dynamically evolve throughout the sequential stages of processing. This allows us to identify and prioritize the structure-property and process-structure links for which we want to build predictive design models and at this point we could use empirical interpolative correlations. However, given the power of the fundamental CALPHAD databases, our commitment from the start was to use mechanistic models parameterized in terms of the fundamental data in CALPHAD to be far more predictive than is possible with empirical modeling. This led to a down selection of tools of computational material science to support this kind of engineering that are represented in Fig. 4. As represented in the bottom of Fig. 4, the biggest impact of quantum mechanical methods has been the prediction of interfacial properties. Next level up, material science has given us the predictive theory of nanoscale precipitation and associated strengthening. At the 3$^{rd}$ level, continuum methods of micromechanics have been applied to the simulation of unit processes of fracture and fatigue. It's this combination of quantum physics, materials science, and continuum mechanics that has given us the capability for performance-driven design. Equally important is the constraint for scale-dependent processability of materials. That's enabled by the materials science models of solid-solid and liquid-solid phase transformations (top of Fig. 4) which are scale dependent through heat transfer. Taking that into account using macroscopic process models for the final scale of production of a material, it's been possible to constrain a design up front to be processable at the final scale of production. The software models and their platforms are represented by the acronyms at the right. Equally important at the left is the advanced instrumentation that has allowed us to calibrate and validate these models and importantly quantify the uncertainty of our model predictions to incorporate in our design practices.

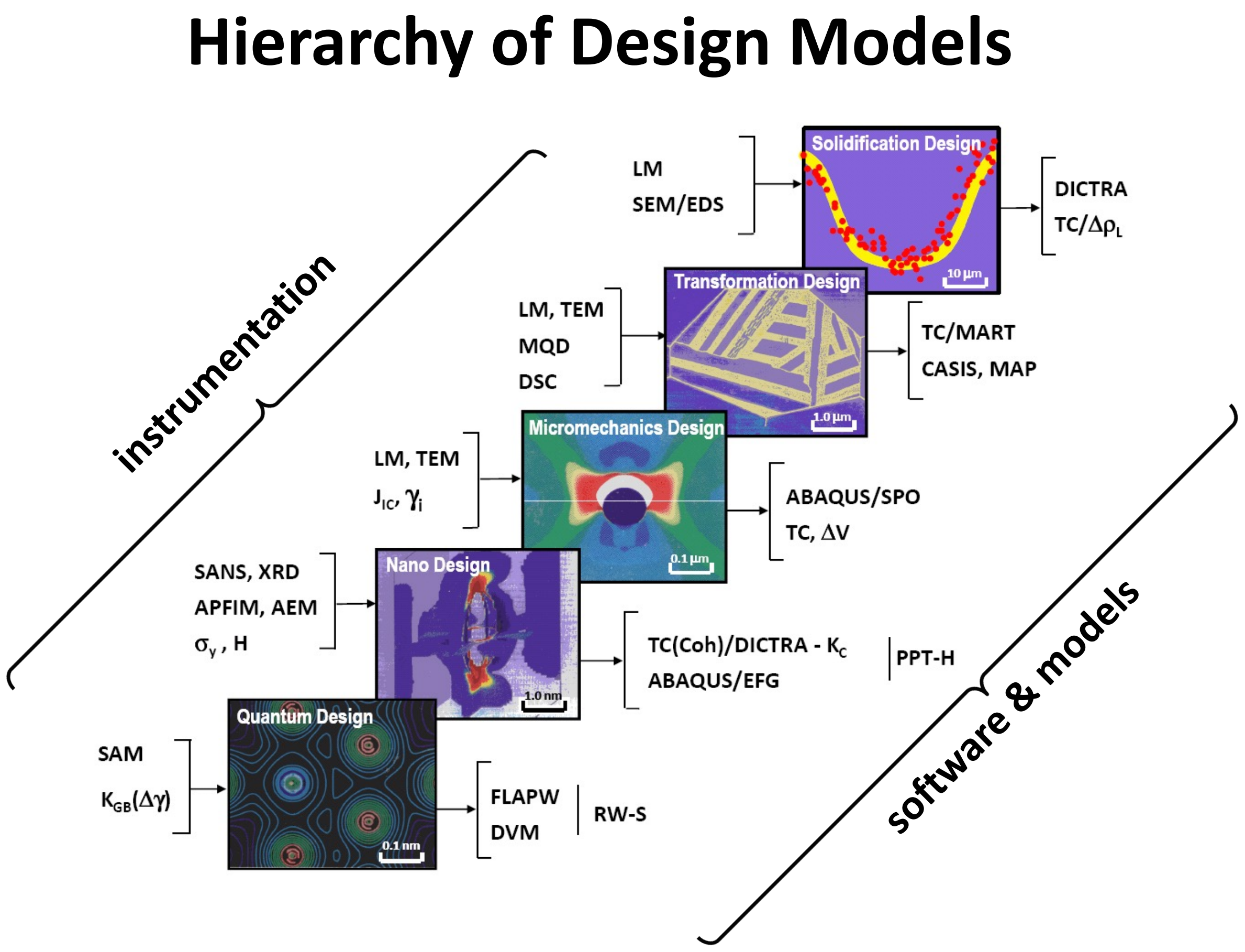

*Fig. 4: Hierarchy of models applied in computational materials design. Acronyms at right represent models and software; acronyms at left represent advanced instrumentation enabling calibration/validation and uncertainty quantification* [8].

### 3.2 QuesTek Innovations

The early research in the 1980s began with the highest performance $M_2C$ carbide secondary hardening steel AF-1410. By a vast combination of experimental techniques, the evolution of those strengthening carbides were fully mapped out. The data validated predictions of the Langer-Schwarz theory of precipitation at high supersaturations where the system evolves by nucleation and coarsening without supercritical growth. Here the length scale is scaled to the initial critical nucleus size which in turn scales inversely with the precipitation driving force, giving us a thermodynamic handle on particle size. It was important to acknowledge though that the trajectory of composition and associated lattice parameters is consistent with coherent nucleation, so it was necessary to add to the chemical thermodynamics the additional elastic energies of the coherent state. With those insights we were able to build a thermodynamics-based approach to strengthening efficiency. That's summarized by the derivation of Fig. 5 starting with the Orowan precipitation strengthening contribution and predicting that it should not only scale with phase fraction to the one-half power but also directly with the magnitude of the precipitation driving force [12].

# Thermodynamics of Strength

$$\Delta\tau_{Or} = K_1 \frac{\mu b}{\lambda} \qquad \lambda \propto \frac{d}{f^{1/2}}$$

$$\Delta\tau_{Or} = K_2 \frac{f^{1/2}}{d} \qquad d \propto |\Delta G|^{-1}$$

$$\Delta\tau_{Or} = K_3 \, |\Delta G| \, f^{1/2}$$

*Fig. 5: Thermodynamic parameterization of dispersion strengthening based on relation of Orowan strengthening to particle size and volume fraction, and scaling of particle size to critical nucleus size* [12].

A series of experimental cobalt-nickel steels were made with a fixed carbon content and thereby fixed final phase fraction that validated this prediction. It was important as well to not only treat the elastic energies of coherency but also acknowledge that the starting state before alloy carbide precipitation is a state of paraequilibrium with cementite that reduces the carbon potential. By taking that into account we were able to show that there's a strong enough driving force dependence for the hardness to vary from the Rockwell C40s to the Rockwell C60 showing a very strong primary contribution [12]. This has then given us the thermodynamic handle that has allowed us to design sufficient strengthening efficiency that, for a given carbon content, we could get 50 percent more strength than previous steels.

As mentioned above, another major advance very early on in the program was applying the all-electron accurate DFT calculations to the surface thermodynamics of interfacial embrittlement to enhance boundary cohesion to offset hydrogen interaction. The framework for that was the Rice Wang model [13] that predicted that embrittlement potency would scale with the energy difference of a solute in the grain boundary and free surface environments. Fig. 6 represents the measured embrittlement potency of phosphorus and sulfur, and the cohesion enhancing contributions of boron and carbon, which represent the most well-studied components in steels. The figure shows the DFT prediction of those energy differences lining up quite well. We were thus able to validate that the predicted thermodynamics and the framework of the Rice Wang model are capable of describing the measured behaviors [12], and we then went on to the much less studied area of the role of the substitutional alloying elements [14].

# Grain Boundary Embrittlement

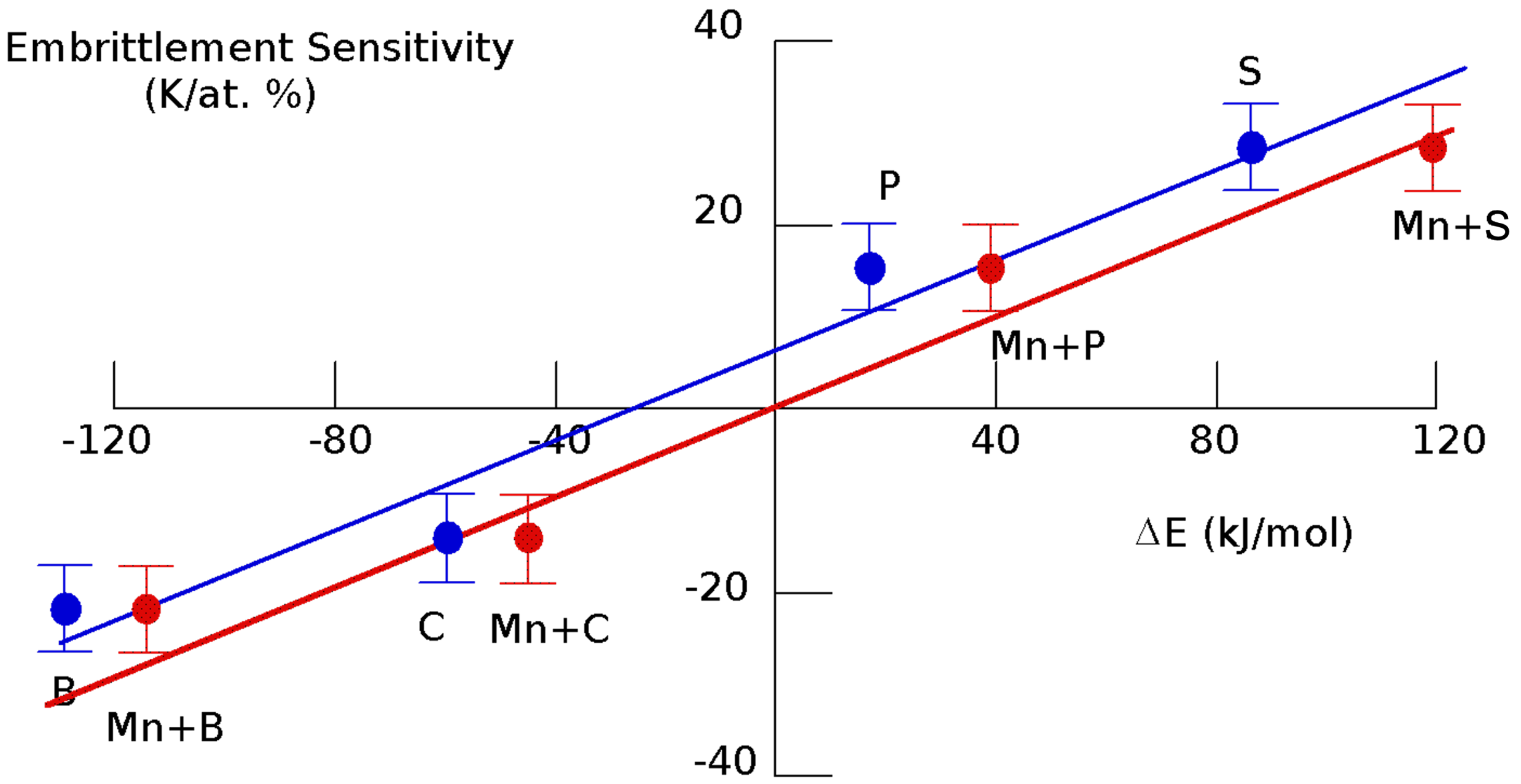

*Fig. 6: Correlation of measured GB segregant embrittlement potency to computed segregation energy difference between GB and free surface environments* [13].

This began with the red points of Fig. 7, predicting this energy difference for Manganese Molybdenum and Palladium [15]. We then used that as calibration points for a data-mining exercise that mapped these quantities back to handbook values of atomic size, individual cohesive energies, and other thermodynamic parameters, to make the prediction that's shown by the black points. This then allowed us to identify the most potent cohesion enhancers, which was followed by the rigorous calculations shown in blue that validated that the simplified model had correctly identified the strongest enhancers. We find that while the rigorous calculations have an accuracy within 0.1 eV, the simplified model was within 0.2 eV which is more than enough to confidently predict the greater than 1 eV potent cohesion enhancers. This type of database was used to then design grain boundary composition in the high-performance steels. We have succeeded in these very high strength landing gear steels to completely eliminate the intergranular form of stress corrosion cracking as a result of this enhanced boundary cohesion. That's one example of a surface thermodynamic genome coming entirely from the DFT calculations.

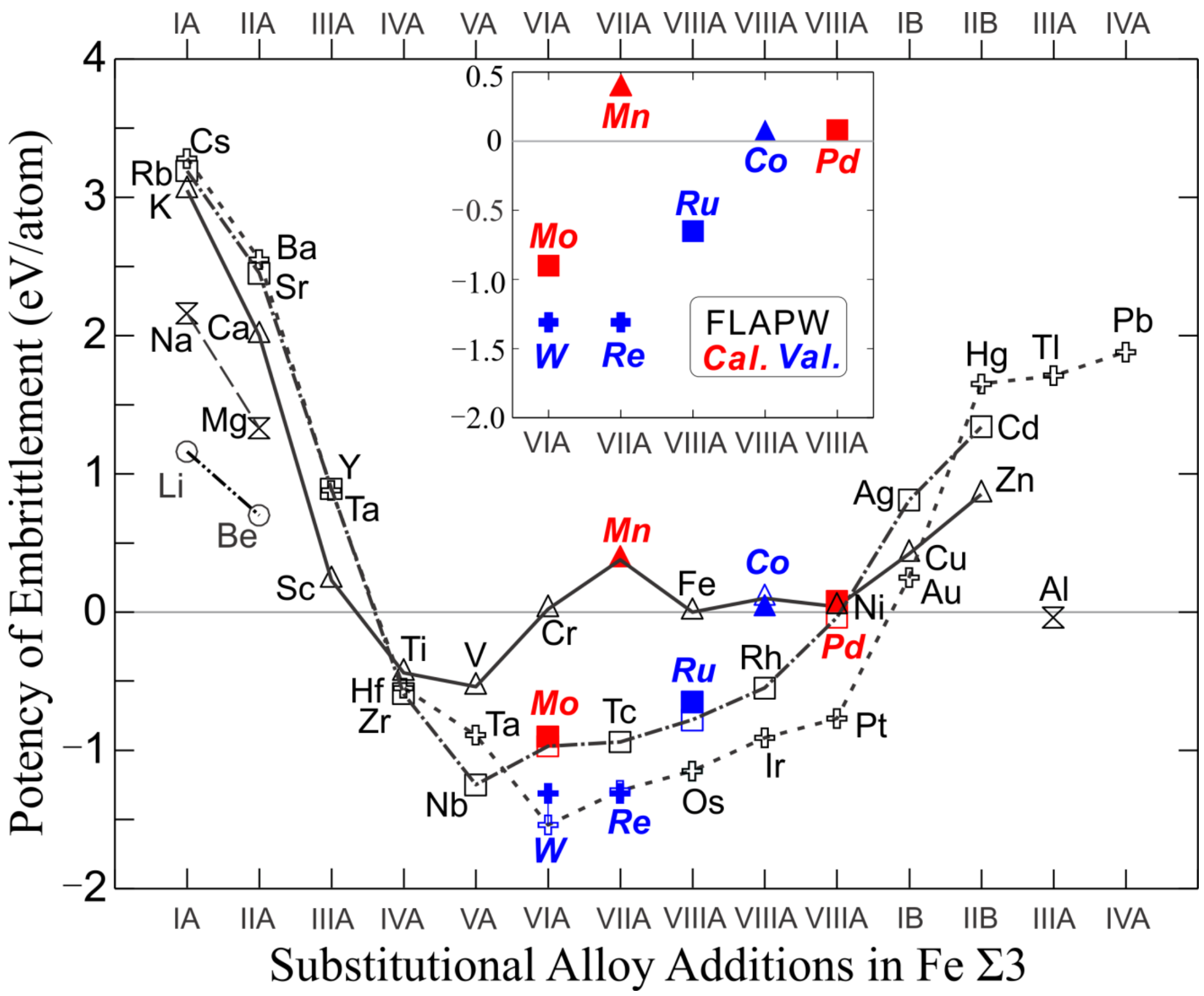

*Fig. 7: Predicted embrittlement potency of substitutional solutes in Fe, based on rigorous FLAPW calculations (color) and phenomenological extension* [14,15].

The actual parametric design integration is performed graphically as shown in Fig. 8. This represents the example of the stainless landing gear steel, where we're looking here at a pair of composition variables, starting with optimizing the nanoscale structure that is produced in the final stage of processing. Contours represent the combination of the thermodynamic driving force plus phase fraction to predict the composition range that meets the strength requirements. Overlaid on that are the processability constraints of the Ms temperature to get a fully martensitic structure as well as contours of the solution temperature necessary to put those components in solution during austenitizing. With that optimized, we back up to earlier stages of processing. The right figure is an optimization of the MX phase grain refiners, with a composition and processing temperature variables shown. We constrain these carbides to be soluble at homogenization temperatures, and able to precipitate in fine form at hot working temperatures, while maintaining a necessary grain pinning phase fraction at the final austenitizing temperature. Similarly, we can back up to earlier stages to define the deoxidants and sulfur gettering compounds that ultimately become the primary inclusions. Doing this graphically also gives us a sense of the sensitivities that allow us to take robust design strategies.

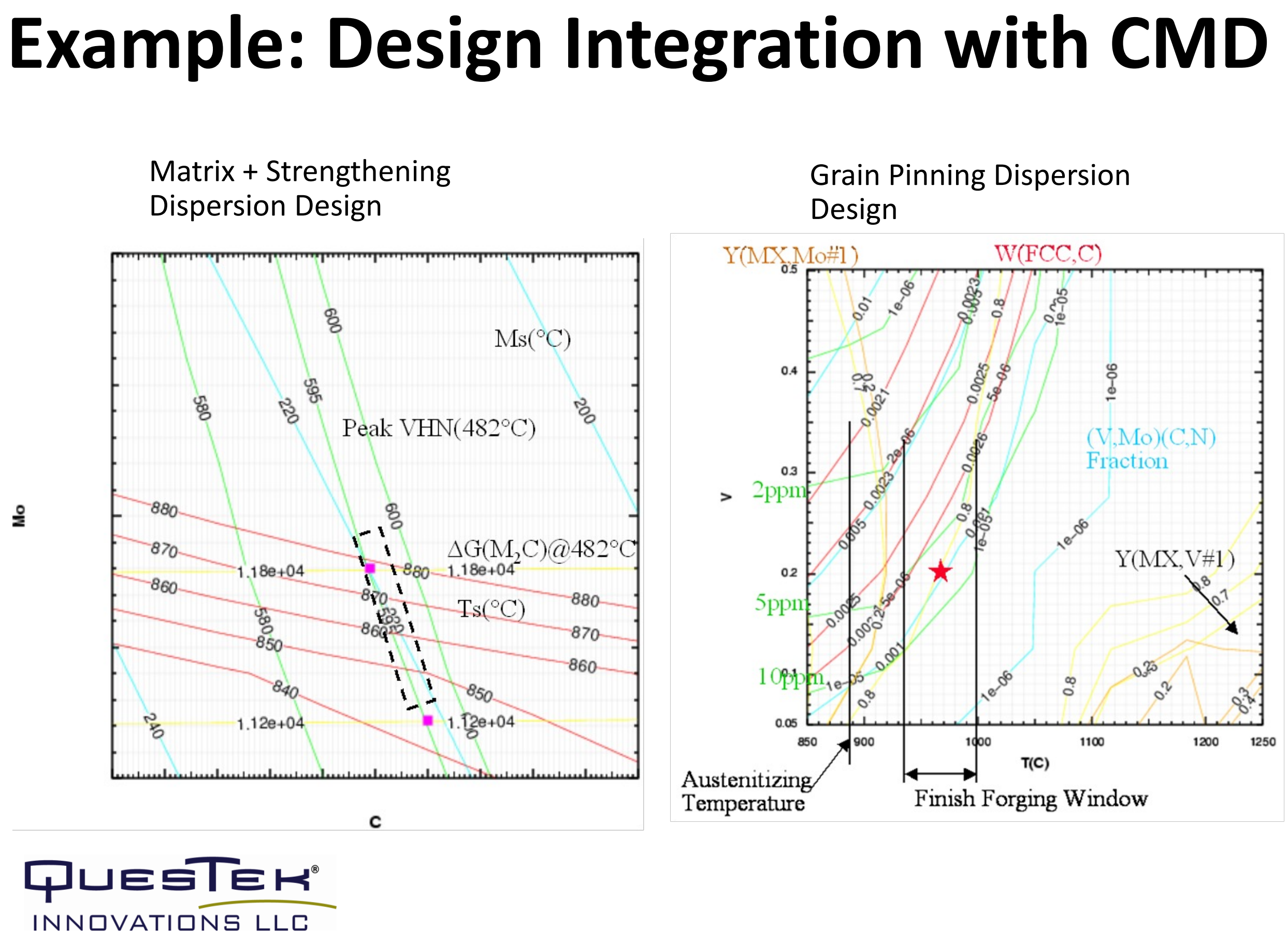

*Fig. 8: Graphical parametric design optimization of nanoscale strengthening dispersion and grain refining dispersion in stainless landing gear steel using CMD design interface* [15].

The first four commercial products to come out of this (Fig. 9) are a pair of high-performance carburizing gear steels that are now being qualified for helicopter applications. We now have two flight qualified aircraft landing gear steels. All four of those designs use the thermodynamics-based optimization of strengthening efficiency that brings the carbide size down to an optimum three nanometers as validated by the early 3D atom probe reconstruction shown at the center of the figure. Efficient strengthening, allowing us to get to the desired strength with less carbon, is helping us to better manage the trade-off with other properties such as fracture toughness or corrosion resistance, enabling each of these steels to be best in class.

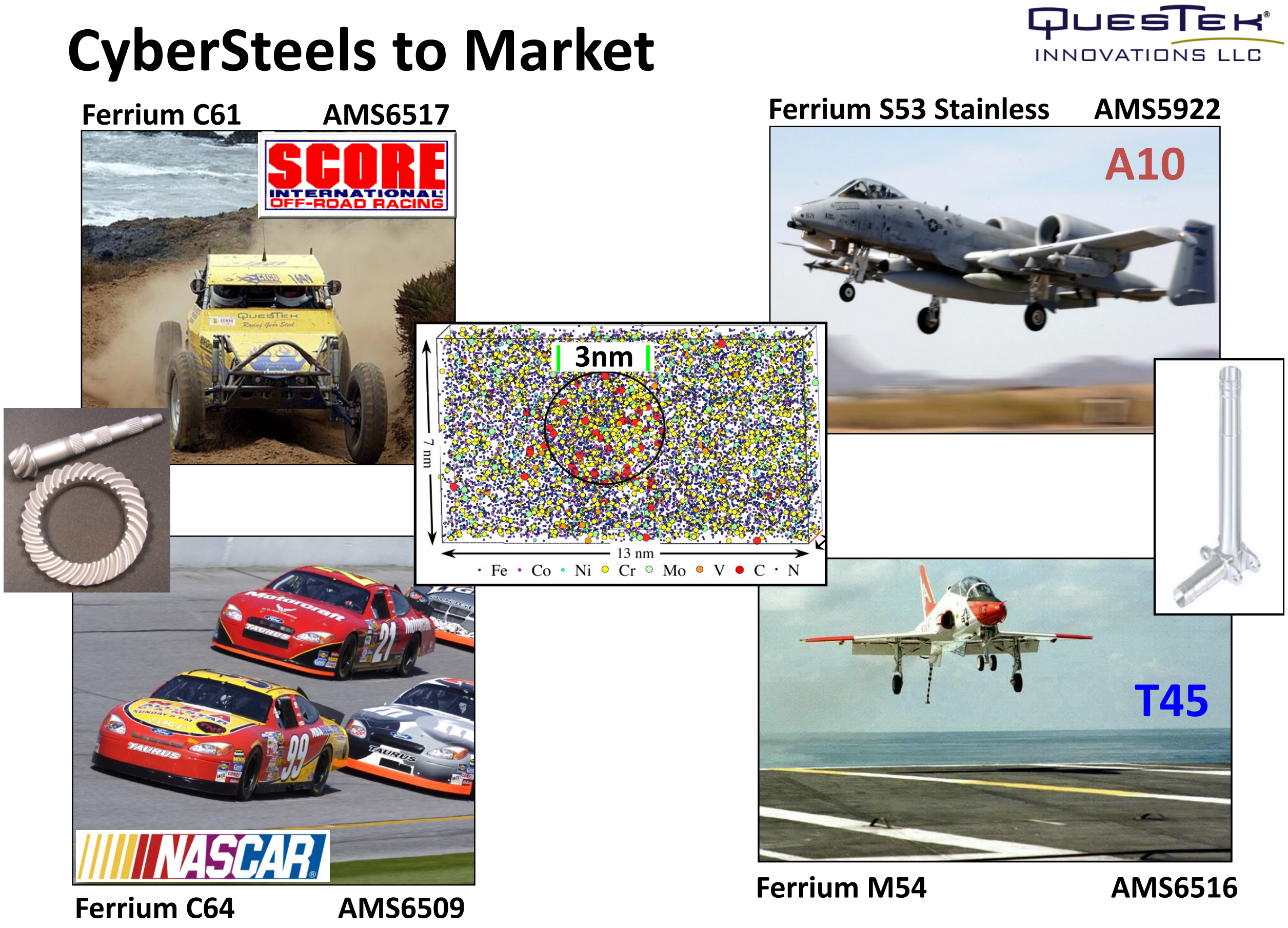

*Fig. 9: First 4 commercial high-performance steels created by CALPHAD-genomic design. All exploit driving-force-based refinement of strengthening carbides to optimum 3nm size, validated by tomographic atom-probe microanalysis* [15].

### 3.3 DARPA AIM Project

The DARPA AIM project (Fig. 10) is where the microstructure models were first coupled to macroscopic process models in the context of process optimization of a nickel-based superalloy turbine disk. For this we initially built the PrecipiCalc simulator [16] using a mean field approximation but grounded in the CALPHAD system, which has now become the commercial product TC-PRISMA. That was the necessary tool to meet the goals of the DARPA AIM project, which was first integrated, through the iSight PIDO (Process Integration and Design Optimization) system, with the other tools of macroscopic engineering, and then used to couple to heat transfer simulations to predict the spatial distribution of microstructure and strength in a turbine disk. It was further applied to the process optimization of a subscale disk whose performance was validated by overspin burst testing. We then went on to the most challenging problem of Monte Carlo simulation of manufacturing variation within the tolerances of composition and processing to get the shape of the final distribution of strength at room temperature and elevated temperatures. Compared to standard statistical inference that in this case involved 700 turbine disks being measured to pin down a one percent minimum property, we demonstrated that by accurately predicting the shape of the distribution we could then calibrate by linear transformation with as few as 15 data points to accurately predict the one percent minimum property within 1 ksi [15]. The methodology proven in the case of the turbine disks was then applied to our landing gear steels, which were the first examples of the application of this methodology to accelerate the qualification of new alloys [7].

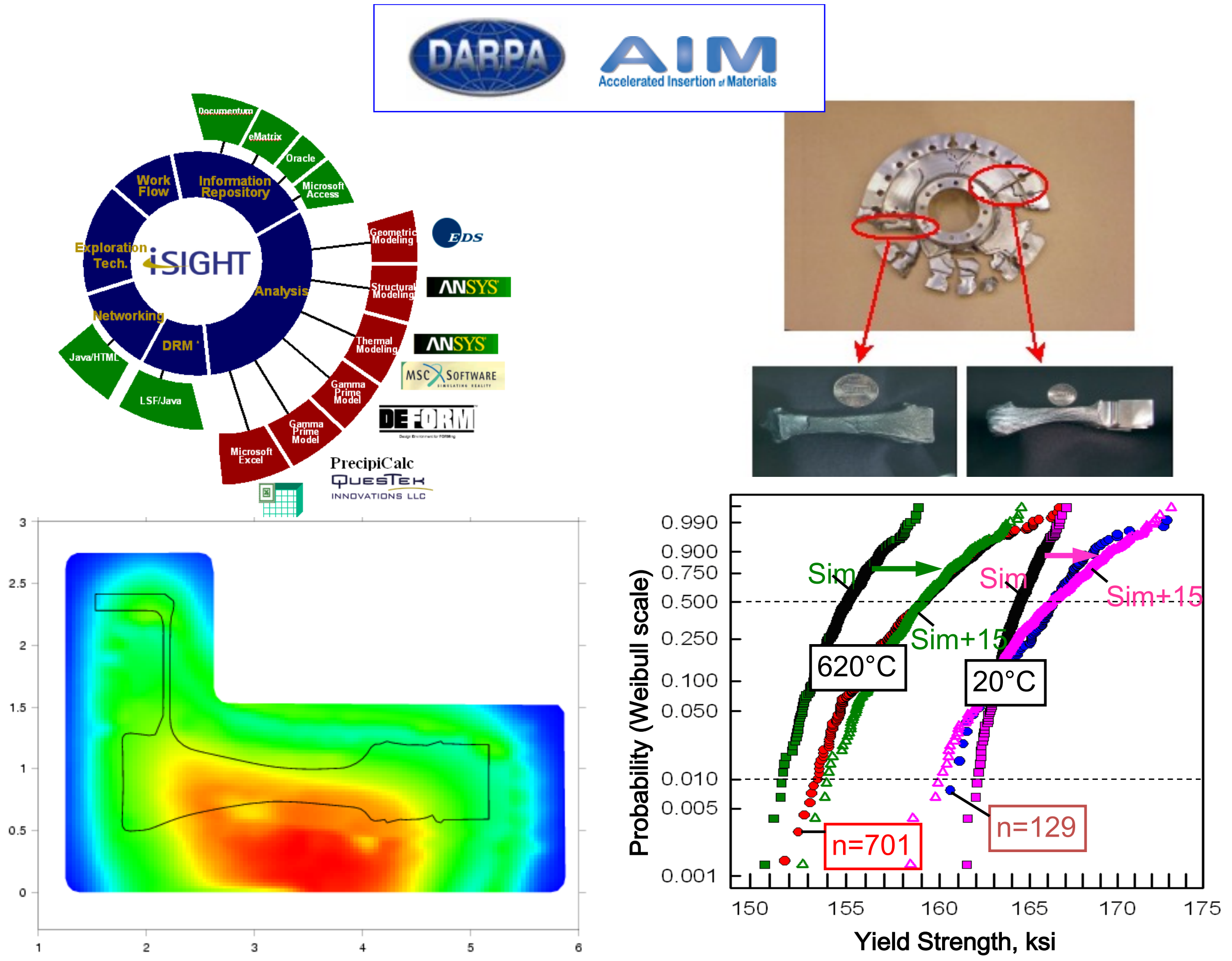

*Fig. 10: Summary of DARPA-AIM accelerated materials qualification technology. Integration by iSIGHT system of tools of macroscopic engineering with PrecipiCalc microstructural simulation, predicting spatial distribution of structure and properties in heat-treated turbine disc forging, correctly predicting disk overspin burst performance. Lower right represents predicted shape of manufacturing property distribution, where linear-transformation calibration with only 15 data points accurately predicts 1% minimum property design allowables* [15,16].

### 3.4 NASA UM Protocol and Design with Uncertainty

A follow-on to DARPA AIM was a four-year project supported by NASA Glenn to go back over the AIM methodology and develop the protocols (Fig. 11) to manage uncertainty within this framework [17]. This was done in the context of process optimization for dual microstructure heat treatment of third generation disc alloys. For each of those new alloys we first brought them to equilibrium and compared the predictions of the $\gamma$ and $\gamma'$ compositions and fractions at equilibrium to down select the most accurate database available. If necessary, to get the exact phase fraction we would apply a rigid energy shift in the GES module of Thermo-Calc to get very accurate phase fractions. After pinning down the thermodynamics we ran diffusion couples against pure nickel and compared the mobility databases and, when necessary, applied a uniform scaling factor to the diffusivity matrix to get high accuracy in our diffusivity predictions. With those pinned, we employed a simple quenched-pin experiment to measure the nucleation undercooling to define the coherent interfacial energy. With that calibrated, we made predictions of the type of slow cooling experiment that would drive the system entirely into supercritical growth, so that we could also establish the interfacial mobility pre-factor for the growth of incoherent primary particles. The final step was the fine tuning of molar volume databases to predict misfit and the corresponding critical size for coherency transition. In this way, we were able to get very high accuracy predictions throughout complex heat treatment. This was then used by Rolls-Royce Indianapolis in the process optimization of their RR-1000 dual microstructure heat-treated turbine discs, a significant achievement of the ICME technology pioneered by the DARPA-AIM program.

# Uncertainty Management:
# NASA Calibration and Validation Protocol

| Experiments | CALPHAD Fundamental Databases | Material Kinetic Model Parameters |
|---|---|---|
| Equilibrium Age + APT and EDS Compositions | Thermodynamics, ΔE | |
| Diffusion Couple + Microanalysis | Mobility, $D_{scale}$ | |
| SSDTA + APT | | $\sigma_{coh}$ , $G_{el}$ (est.) |
| Coarsening Age, Slow Cool + SEM/TEM for $\gamma'$ size and fraction | | $\sigma_{incoh}$, $M_0$ |
| XRD, TEM for misfit | Molar Volume | $G_{el}$, $R_{coh \rightarrow incoh}$ |

QuesTek INNOVATIONS LLC

*Fig. 11: Sequential calibration of thermodynamics, atomic mobilities, interfacial properties, and molar volumes in support of high-accuracy optimization of complex turbine-disc thermal processing* [17].

The uncertainty management protocol developed under the NASA Glen project has had an important impact on the actual "design with uncertainty" practices that we've then established. It was in the late 90s that Howmet called to our attention the problem of scaling up the processing of single crystal turbine blade alloys from the aero turbine scale to the very large scale of industrial gas turbines. The problem here is the macrosegregation defects, so-called "freckles" that are driven by thermosolutal convection. The physics of the problem was already well established in the late 90s. We took a look at the ability to predict the critical Rayleigh number for this by predicting the solid/liquid partitioning during solidification and the associated liquid molar volumes in order to predict the buoyancy that would drive this process. It was clear in the late 90s that CALPHAD wasn't ready. Our databases did not have the accuracy to sufficiently control those parameters. Fortunately, 15 years later we got the opportunity at QuesTek to come back to this problem under DOE support [18]. We took the multi-database strategy and found the best database for each attribute. Now we could not only constrain the partitioning and buoyancy of the liquid to be processable, but also were able to use the multi-component thermodynamics to more efficiently partition Rhenium between $\gamma$ and $\gamma'$ and make more efficient use of this expensive element. This resulted in a 1 wt.% Re single crystal alloy that had the processability of a Re-free alloy and could go all the way to IGT scale as demonstrated in collaboration with Siemens. With the efficient Re partitioning, this 1 wt.% Re alloy gave the creep performance of a 3% Re alloy as summarized in Fig. 12. It was a victory for the use of uncertainty management within this CALPHAD framework to solve a very complex sophisticated problem.

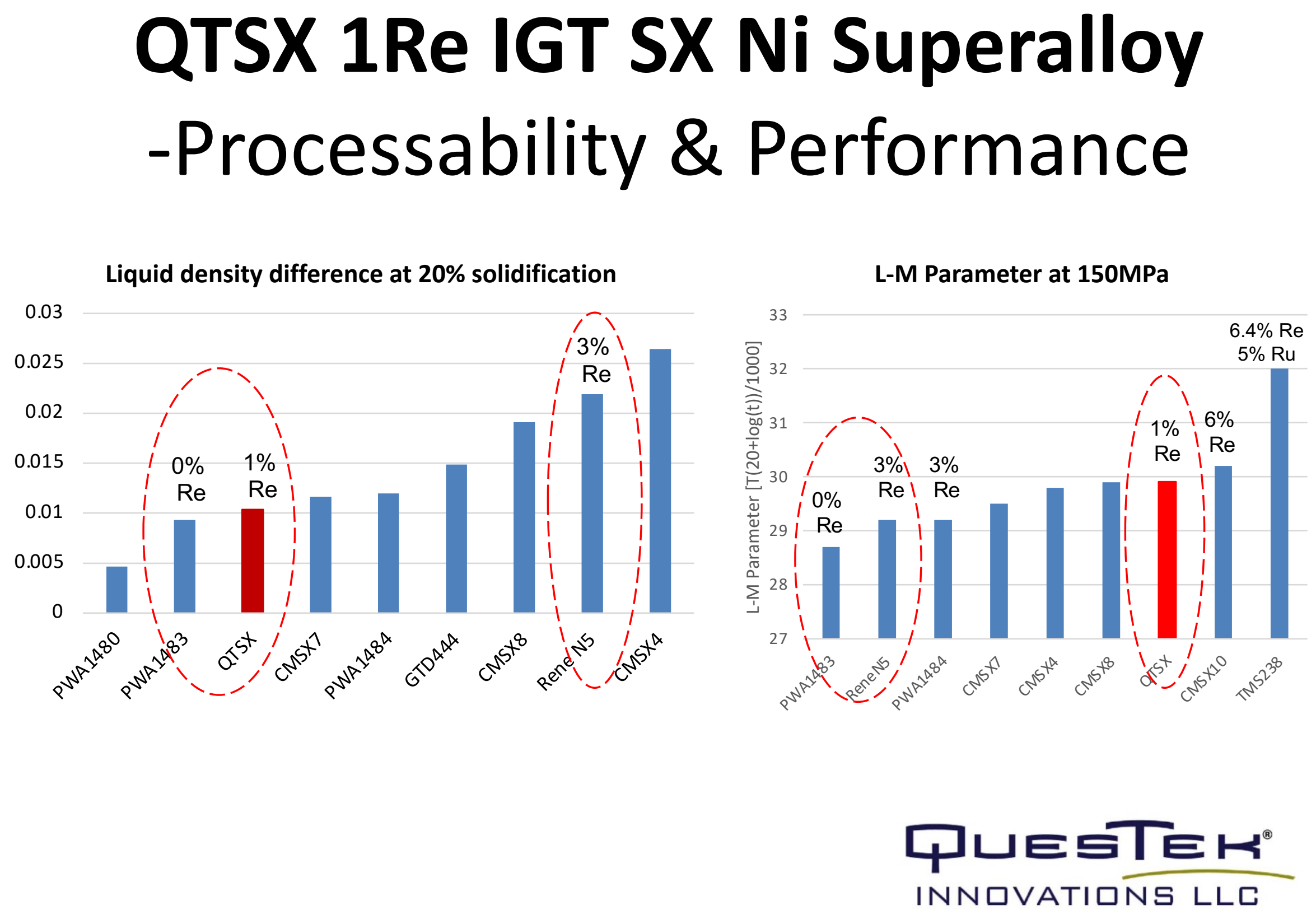

*Fig. 12: Scalable 1Re single-crystal Ni superalloy designed with multi-database strategy to combine processability of Re-free alloy with creep performance of 3Re alloy* [18].

### 3.5 CHiMaD and Additive Manufacturing

The context of current university research now is the multi-institutional "CHiMaD" Center for Hierarchical Materials Design that functions as the lead research center of the MGI. Within this center, a subgroup focuses on the design of new alloys for additive manufacturing. Starting with the DARPA "Open Manufacturing" program, QuesTek at this point has had over 60 projects in this area, and provides a foundation for this effort. Projects address not only the design of new alloys that can get more value out of the additive manufacturing processes, but also the adaptation of the AIM methodology to very efficiently predict location specific minimum properties in printed components to accelerate their qualification. Along the way to constrain materials design for resistance to cracking during printing, we developed a deposition cracking resistance parameter that employs the "CSC" parameter developed in casting technology for resistance to solidification hot tearing, and combines that with the freezing range, coming from the welding technology to constrain against liquation cracking. Combining this with a third criterion on the quench suppressibility of precipitation strengthening, to avoid "strain-age" cracking, has provided the constraints to design specifically for the additive processing. High performance can then be achieved by exploiting the unique characteristics of printing, particular laser fusion processes, where we combine boiling melts with very rapid solidification [19].

Our best example of designing to achieve microstructures that could not be achieved by wrought processing is a $L1_2$ strengthened high temperature aluminum. As shown by the equilibrium step diagram of Fig. 13, in the ground state structure we exploit an aluminum nickel eutectic for the hot tearing resistance, combined with a dispersion of a multi-component $L1_2$ phase [20]. Not present in the equilibrium diagram is another very important constraint that under non-equilibrium Scheil solidification the system will form a metastable ternary phase which becomes the reservoir for the rare earth elements, which by rapid solidification is finely dispersed. That allows us to support the precipitation at an even finer scale of the $L1_2$ precipitates and maintain high strength at high temperatures (outperforming the scandium-bearing Scalmalloy) with a scandium-free $L1_2$ with good strength in the region of 250 to 300 C (Fig. 14) [20]. This represents a unique capability enabled by a unique microstructure by making optimal use of the unique process pathway of laser-fusion additive manufacturing.

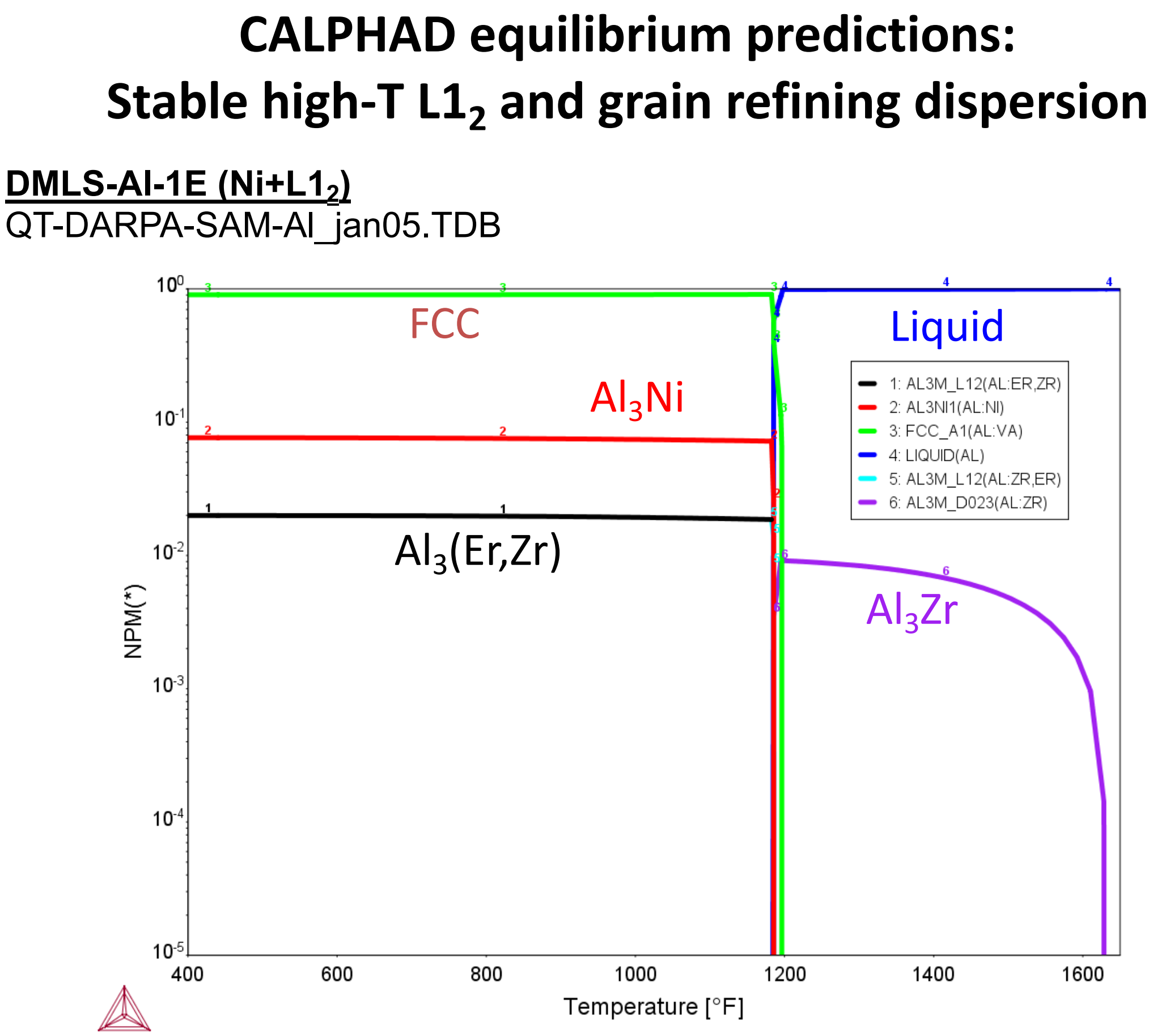

*Fig. 13: Equilibrium step diagram of printable alloy exploiting stable L1$_2$ dispersion for high-temperature strength* [20].

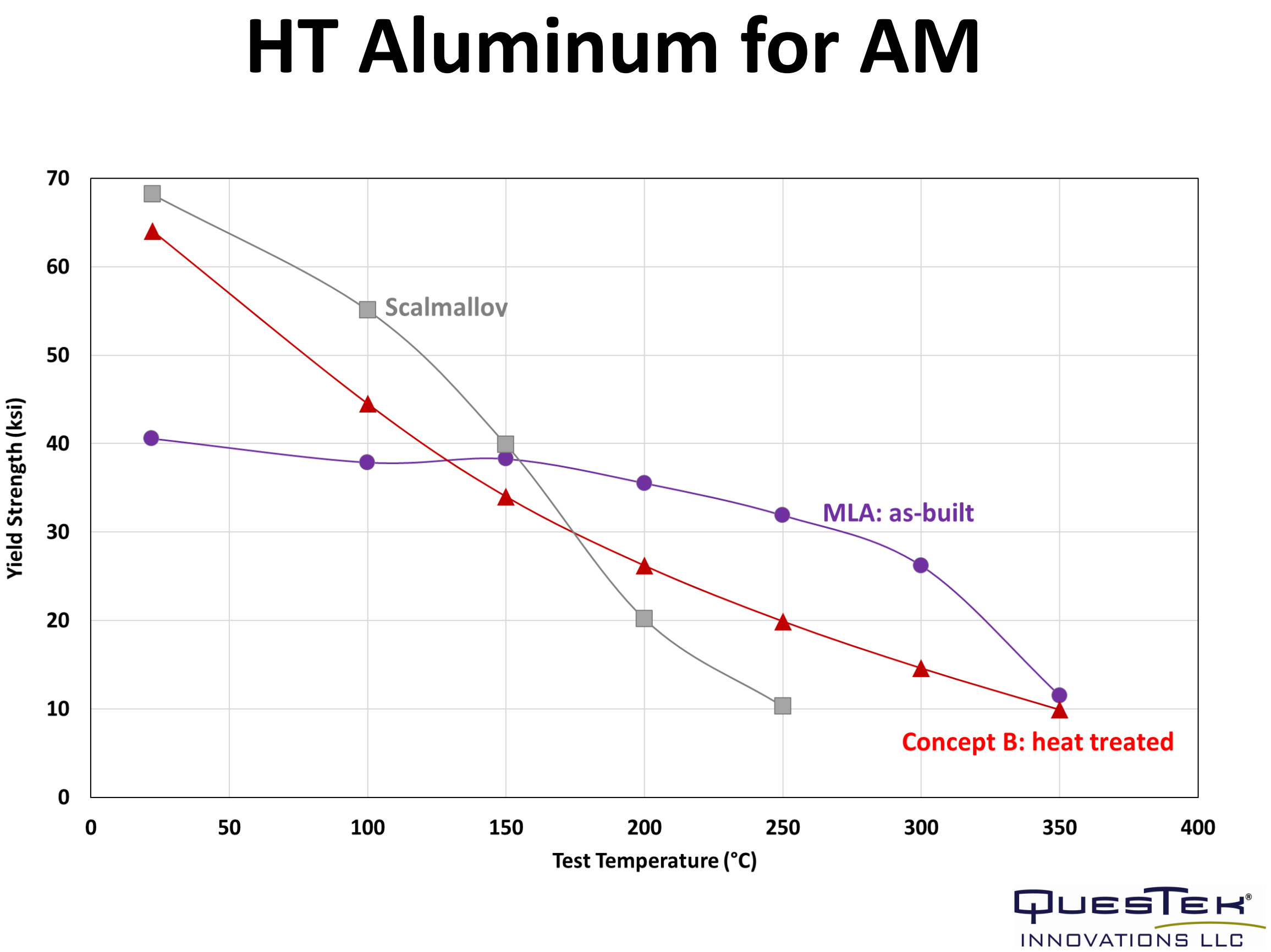

*Fig. 14: Superior high-temperature strength retention of printable L1₂-strengthened aluminum alloys* [20].

We next briefly return to the original goal of Kaufman and Cohen in using accurate low temperature CALPHAD thermodynamics to get at the mechanism of martensitic transformations. Eddie Pang, a doctoral student advised by Prof. Chris Schuh, volunteered to coach a project in our first offering of the Computational Materials Design class at MIT in 2020. The resulting feasibility demonstration helped redirect his thesis towards CALPHAD-based design. Quantitative applications of the kinetic theory of martensitic transformation [10] delivered a cracking-resistant $ZrO_2$-based shape memory ceramic with record low transformation hysteresis.

Eddie received the best doctoral thesis award in the materials department based on that work, with publication in the journal Nature [21].

## 4 Advances in the Development of CALPHAD Databases

As mentioned above, Kaufman and Cohen laid the foundation of CALPHAD modeling in the iron-nickel system where the free energy of the bcc-Ni was needed in order to define the free energy of the bcc solid solutions so the driving force for the formation of martensite could be evaluated [9]. The term, “lattice stability”, was coined by Kaufman for this very purpose [22] and was a hotly-debated topic [23] in the community. In order to efficiently develop multicomponent thermodynamic databases through international collaborations, it was necessary for everyone to use the same “lattice stability” for every element so CALPHAD modeling performed by individual groups could be combined. Several versions of lattice stability were developed with the currently used one published in 1991 [24], and efforts in comparing and integrating DFT predicted lattice stability with the CALPHAD lattice stability are ongoing [25].

In the CALPHAD modeling, any modification of a constitutive subsystem, including pure element, has a compounding effect on the description of a multicomponent system because it affects every description of systems that includes this subsystem [26]. Fig. 15 represents this compounding effect illustrating the number of binary and ternary re-assessments needed if one unary description is changed in a six-component system [26]. This challenge prompted Liu’s group to develop the high throughput computational tools for efficient CALPHAD modeling supported by DFT-based first-principles calculations and deep neural network machine learning models. Liu’s group started to develop an automated CALPHAD modeling tool, Extensible

Self-optimizing Phase Equilibria Infrastructure (ESPEI), demonstrating its promises [27]. Recently, the group developed an open-source software package, PyCalphad, for thermodynamic calculations [28] and used it to develop a complete new ESPEI code [29], which are being developed as part of an open source software ecosystem, introduced through workshops organized by the nonprofit Materials Genome Foundation, and used in academia and commercial communities notably including SpaceX.

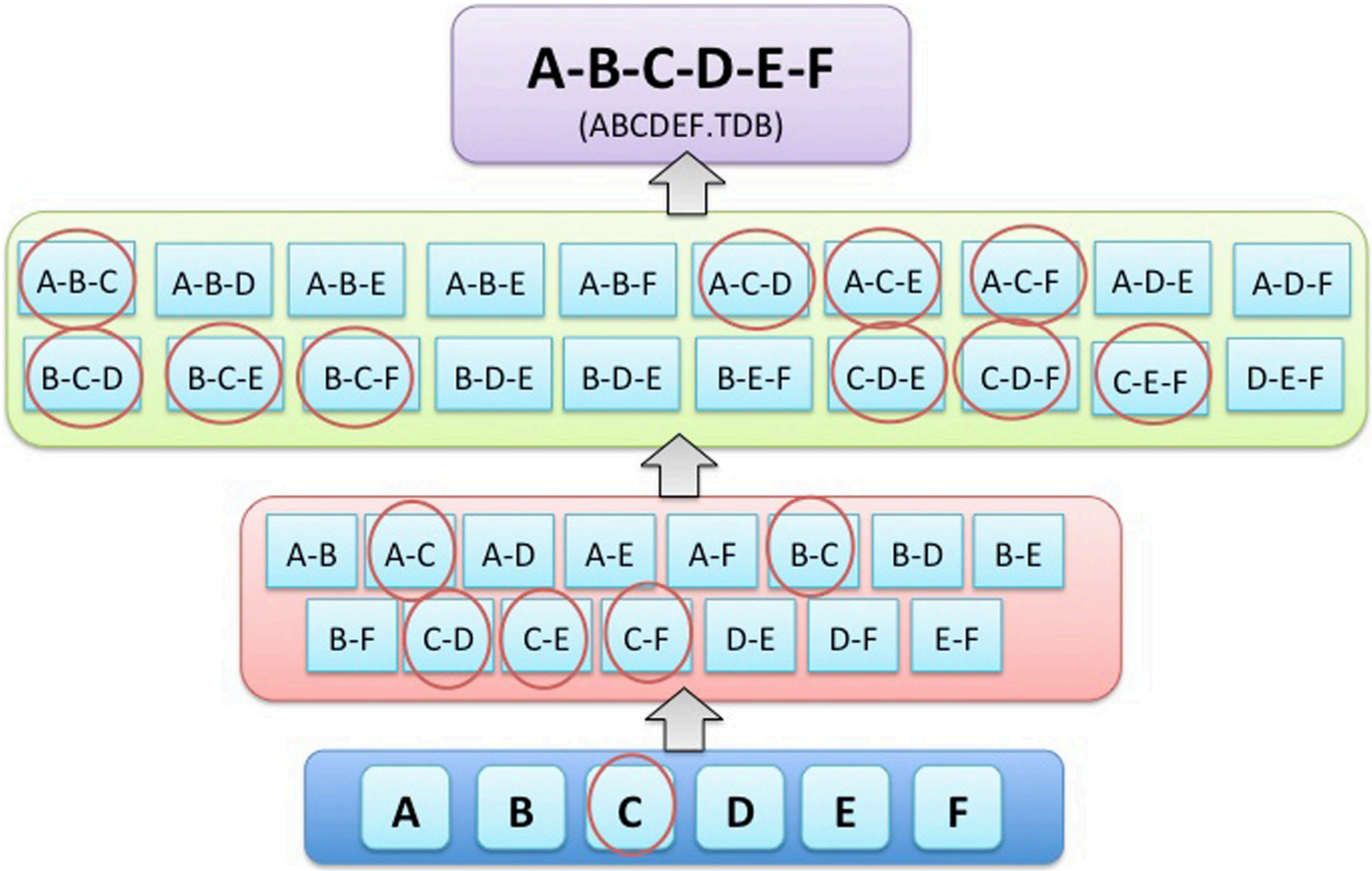

*Fig. 15: Data dependencies in the development of multicomponent CALPHAD databases illustrated for a six component A-B-C-D-E-F system that if the component C is changed, all the circled binary and ternary systems need to be re-evaluated* [26].

As an open-source Python library, PyCalphad [28] supports designing thermodynamic models, calculating phase diagrams and investigating phase equilibria using the CALPHAD method and can read thermodynamic databases and solve multicomponent, multi-phase Gibbs energy minimization problems. A unique feature of PyCalphad is that the thermodynamic models of individual phases are internally decoupled from the equilibrium solver, and the models themselves are represented symbolically, such as the modified quasichemical model in quadruplet approximation (MQMQA) recently implemented [30]. Consequently, the databases can be programmatically manipulated and overridden at run-time without modifying any internal solver or calculation code. The popular Git source code control (SCC) system is used to manage the source of PyCalphad with a suite of continuous integration (CI) tests designed to verify that a revision to the code does not cause unintended behavior, which are run automatically every time a new revision is pushed to the Git repository on GitHub. The rigor and impacts of PyCalphad enabled its winning of the runner-up (2$^{nd}$ place) in NASA Software of the Year competition in 2019 [31] with strong support letters from domestic and international users in industry, national laboratories, and universities.

ESPEI implements two steps of model parameter evaluation: generation and Markov Chain Monte Carlo (MCMC) optimization [29]. The parameter generation step uses experimental and first-principles data describing the derivatives of the Gibbs energy to parameterize the Gibbs energy of each individual phase, resulting in a complete thermodynamic database from those derivatives, commonly called thermochemical data. As phase equilibria require more accurate descriptions of Gibbs energies of all phases, the parameters of all phases are optimized iteratively to be self-consistent. ESPEI uses MCMC to perform a Bayesian optimization of all model parameters simultaneously through an ensemble sampler algorithm implemented in the emcee package with parallelizable ensemble samplers. Three types of data are defined in ESPEI: single phase thermochemical data of the temperature derivatives of the Gibbs energy, activity data (converted into chemical potential) related to the composition derivatives of the Gibbs energy, and multi-phase equilibria data. Their errors are assumed to follow a normal distribution with default values for the standard deviations of each type of data provided in ESPEI, and users can modify the values by adding a weight for each type of data or for each individual dataset. Parameter uncertainty is then evaluated within the MCMC optimization step of ESPEI by quantifying the distribution of the parameter values that make up each converged Markov chain, enabling the uncertainty quantification in terms of models, model parameters, and model predictions [32].

Experimental thermochemical data are typically too sparse to fully describe the Gibbs energies of the phases. The community has utilized DFT-based first-principles calculations to obtain formation energy at zero K and finite temperatures [33]. Liu's group recently developed a high throughput DFT Tool Kits (DFTTK) to automate the calculations of free energy of any given configurations [34] and a mixed-space approach with the short-range interactions accounted for by supercells in real space, the analytical solution for the origin in the reciprocal space which represents the infinite in real space, and an interpolation scheme between them [35]. To further increase the efficiency to generate energetics of large numbers of configurations, Liu's group developed deep neural network machine learning models, named SIPFENN (Structure-Informed Prediction of Formation Energy), for predicting formation energy at zero K [36], which can be installed through PyPI by "pip install pysipfenn" and is being extended for prediction of free energy. During this process, a MongoDB database with over 4 million relaxed structures from DFT-based calculations and experiments [37] was established and used to predict formation energies and crystal structures of new phases [38].

A schematic chart for a data ecosystem including tools to generate, process, use and re-use data is shown in Fig. 16. The complexity in generating proto data is ranked in the ascending order of SIPFENN, DFTTK, *ab initio* molecular dynamics (AIMD), finite element method (FEM), and experiments. The use of data include the inverse design of materials and manufacturing [39] and simulations of materials manufacturing, service, and recycling. The newly generated experimental data from materials manufacturing, service, and recycle feed back into the proto data. Liu's team has recently developed such a data ecosystem for refractory high entropy alloys (RHEAs) [40] and used it for their inverse design [39] as shown in Fig. 17.

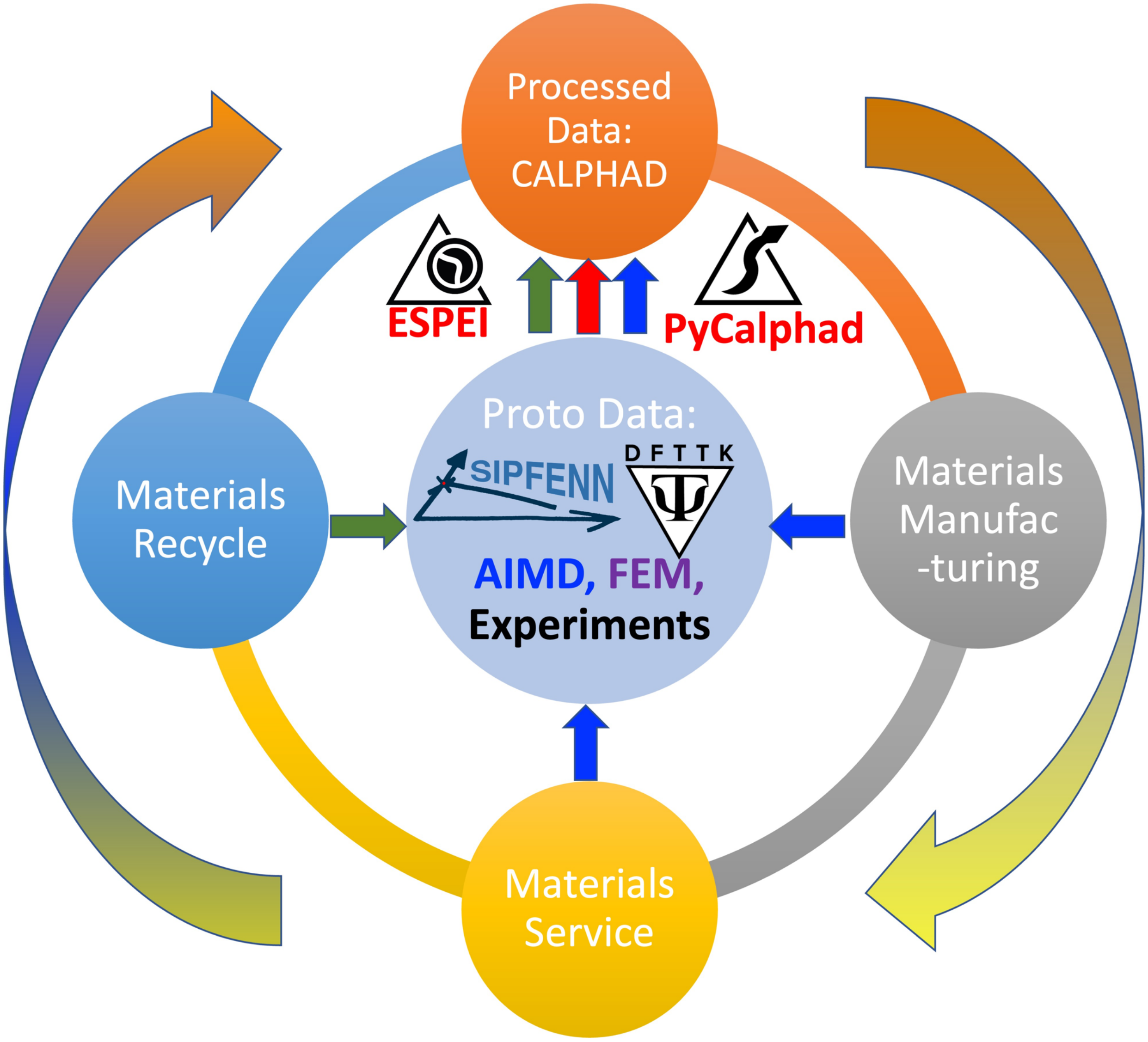

*Fig. 16: Schematic chart of the data ecosystem including proto data (experiments and computations with the latter including machine learning with SIPFENN* [36] *as an example, first-principles calculations with DFTTK* [34] *as an example, ab initio molecular dynamics, and finite element method), processed data (CALPHAD modeling with PyCalphad* [28] *and ESPEI*

[29]*), data for and from materials manufacturing (inverse design of materials* [39]*), and experimental data from materials service and materials recycle.*

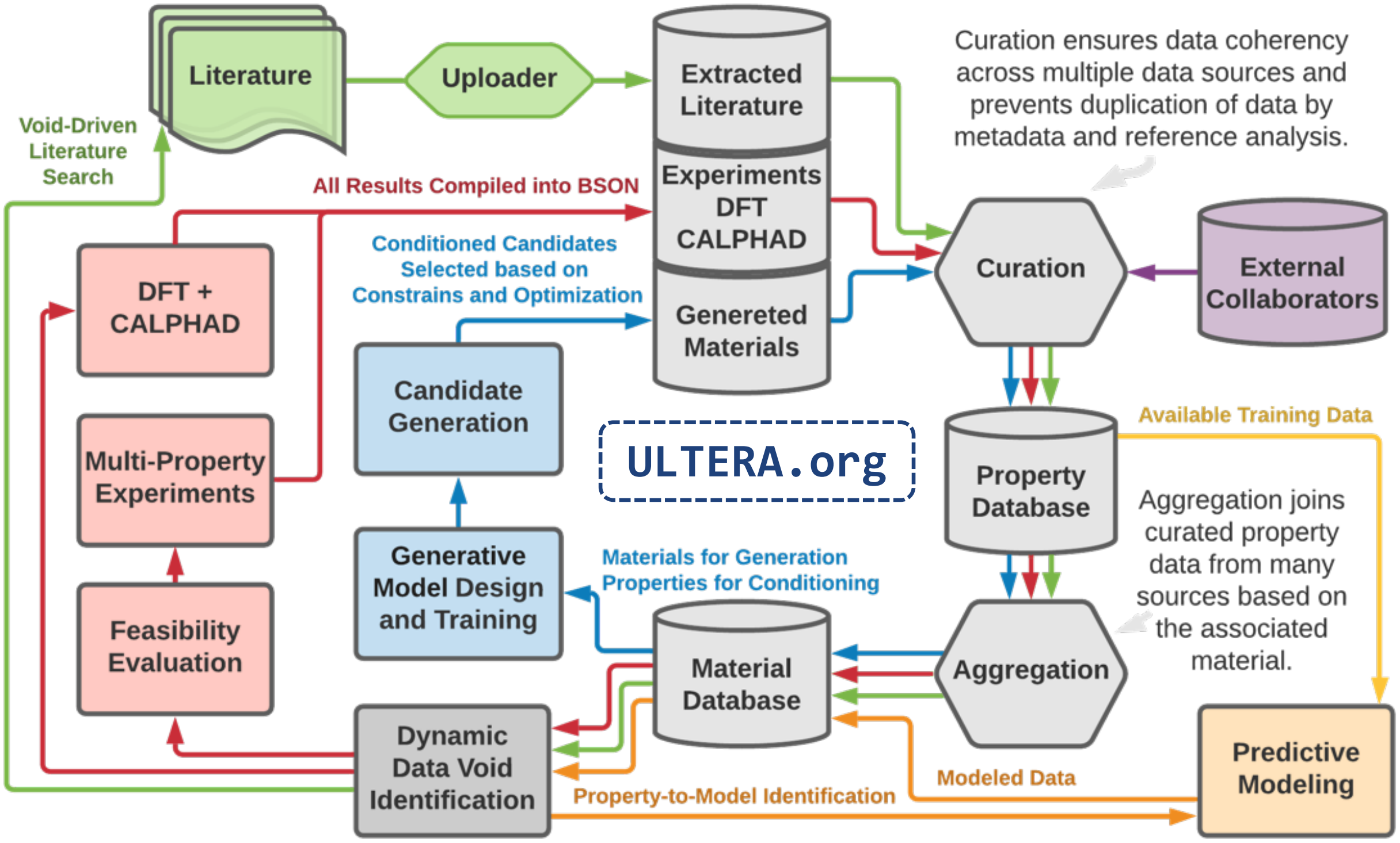

*Fig. 17: Schematic chart of the data flow in ULTERA (ULtrahigh TEmperature Refractory Alloys) data ecosystem with various data sources and types homogenized and combined through curation, then reorganized around unique alloys through aggregation, and finally passed to inverse design (blue), predictive ML modeling (yellow), validation (red), and literature collection (green) routines, closing the cycles of data flow* [40].

DFT-based calculations are often employed for the ground-state configuration of a system, while at the finite temperature, the system consists of statistical fluctuations of non-ground-state configurations in addition to the ground-state configuration. It is thus inevitable that DFT-based predictions, even with thermal electronic and phonon contributions included, differ from experimental observations. To bridge this gap, Liu's group developed the recently termed zentropy theory that integrates DFT and statistical mechanics as shown in Fig. 18 and results in the replacement of the internal energy of each individual configuration by its DFT-predicted free energy [41]. The zentropy theory is capable of accurately predicting the free energy of individual phases, transition temperatures and properties of magnetic materials with inputs of individual configurations solely from DFT-based calculations and without fitting parameters. Those predictions include the singularity at critical points with divergence of physical properties, negative thermal expansion, and strongly correlated physics, and may be applicable to a broad range of transformative properties and systems.

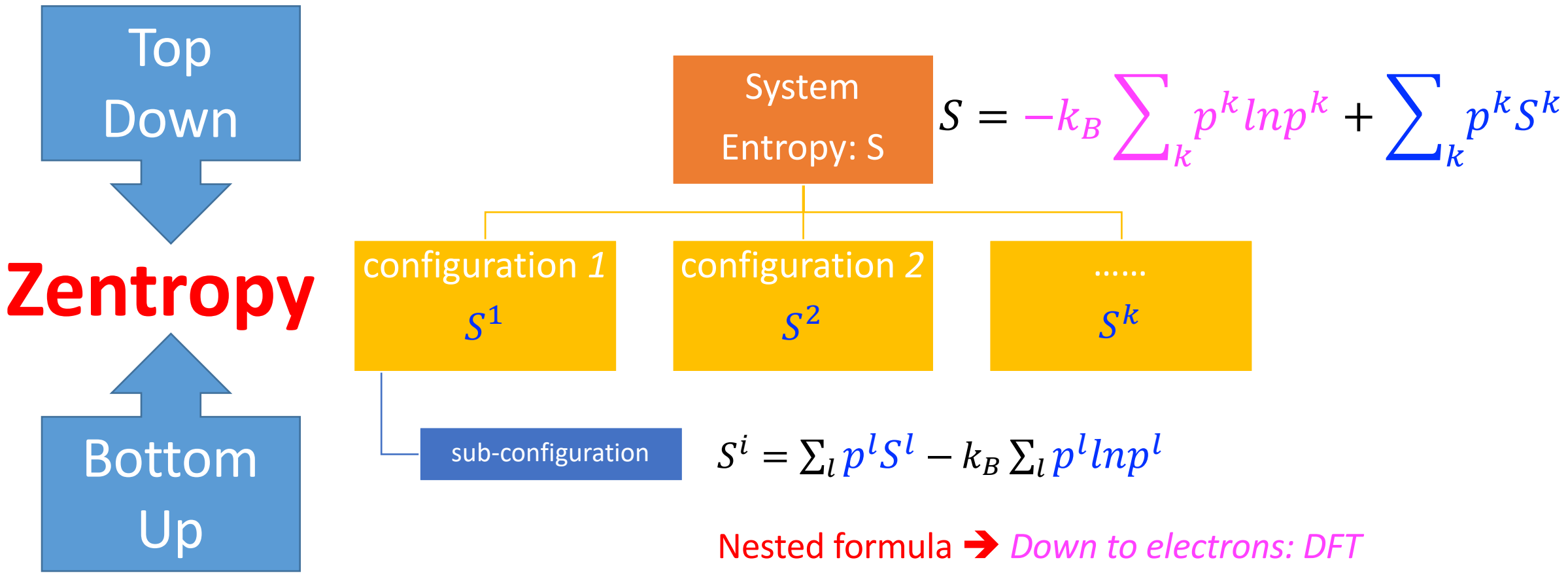

*Fig. 18: Schematics of zentropy theory in its nested formula going all the way down to electronic structures predicted from DFT.*

## 5 Broad Adoption of CALPHAD-based Materials Technology

The CALPHAD based accelerated qualification of QuesTek's second landing gear steel was the subject of a NIST case-study for the MGI that looked at both the technology accelerators and the inhibitors in doing this within a small business environment. It concluded that the technology demonstrated was capable of a three-year cycle; an important benchmark in finally enabling the concept of concurrent engineering denoted at the top right of Fig. 1. The first historic example of that was the four new alloys of the Apple Watch announced in 2014. Those alloys were designed concurrently with the device and delivered in under two years from acquiring the CALPHAD-based technology that enabled it.

Taking concurrency to a new level, it became public in 2015 that Dr. Charlie Kuehmann, after three years at Apple, moved to become Vice President of Materials Technology at both SpaceX and Tesla reporting directly to Elon Musk. This was followed by numerous innovations in materials concurrency heralded by the Chief Tweeter. The tweet of Fig. 19 in 2018 announced the burn-resistant nickel superalloy, designed at QuesTek and fully developed at SpaceX, that serves as the central enabler of the concept of the Raptor Engine for the Mars Starship. The design criterion for the molten-oxide-based flame extinction was actually developed by Zi-Kui Liu in his time at QuesTek before taking a faculty position at Pennsylvania State University. The burn resistance enables the very high oxygen pressures that allow the Raptor Engine its efficiency. Kuehmann has given lectures on how SpaceX has taken the standard target cascade and validation system of concurrent engineering and fully collapsed it into total concurrency. That has very significantly been aided by the integration of materials at an early stage into the concurrency process, taking advantage of the intrinsic reliability and predictability of a designed material. It is a truly new concurrency.

A more recent example is the major scale up of high-performance aluminum die casting to Tesla's "GigaCasting" level, enabled by a radical new die casting alloy delivered in less than 2 years, concurrently with building of the new casting plants, providing the frame of the model Y Tesla. This could well be the greatest innovation in lightweight automotive technology to build cars by this kind of high-scale die casting. Building on this development, it was announced recently that Musk is a strong enough believer in this new technology that, in addition to the two materials groups that Kuehmann has at the two companies, they have now founded a new group at Tesla as the Materials Applications Team to accelerate the adoption of these designed materials. Musk no longer wants to use the legacy alloys of empirical development, but wants all his materials designed, so they are better and more reliable. At this point, Elon Musk may be the most influential advocate for CALPHAD-based materials technology.

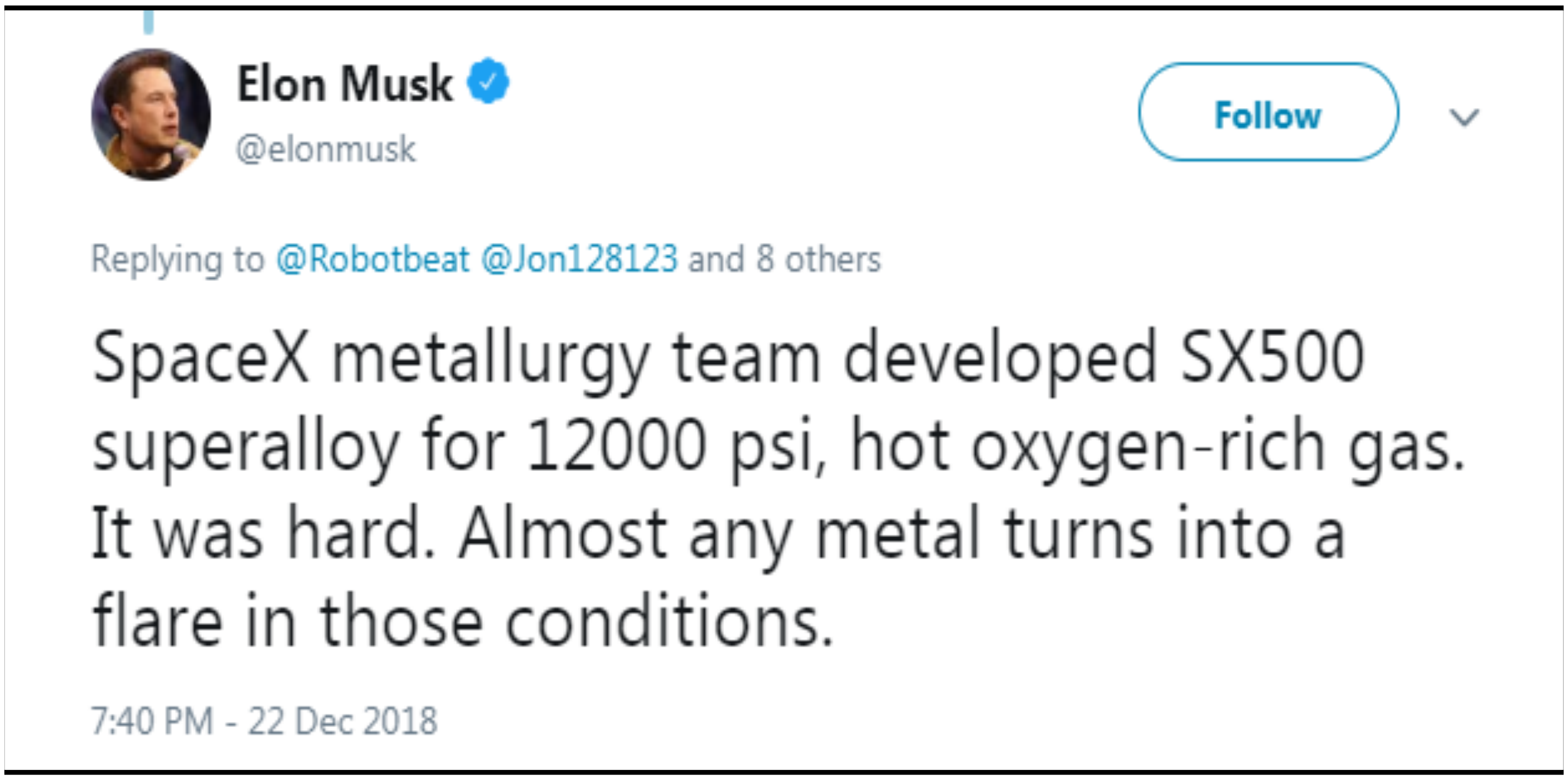

*Fig. 19: Acclaim of computationally-designed, high-pressure-oxygen-enabling, burn-resistant Ni superalloy for SpaceX Raptor engine.*

## 6 Summary

It is a great thing to see these events unfolding. In closing, if we look back to its origins and ahead to its future, the true meaning of CALPHAD is CALculation of PHAse Dynamics.

## 7 Acknowledgements

GBO is grateful to Thermo-Calc Software AB for support of his chair at MIT as Thermo-Calc Professor of the Practice, and to the NIST CHiMaD Design Center and the Office of Naval Research for support of computational materials design research. ZKL acknowledges the support from Dorothy Pate Enright Professorship at Penn State, National Science Foundation (FAIN-2229690, CMMI-2226976, CMMI-2050069), Department of Energy (DE-SC0023185, DE-NE0009288, DE-AR0001435, DE-NE0008945), Office of Naval Research (N00014-21-1-2608), Army Research Lab, Air Force Research Office, National Aeronautics and Space Administration, and many industrial companies, with the current grants showing in parenthesis. ZKL would also like to thank Adam Krajewski for Fig. 18 from the Department of Energy ARPA-E ULTIMATE program (DE-AR0001435).

## 8 Data Availability

There are no new data in the paper.